\documentclass[10pt,letterpaper,twocolumn]{article} 

\usepackage{ol2}
\usepackage[draft]{hyperref}
\usepackage{amsmath}
\usepackage{indentfirst}

\begin{document}

\twocolumn[ 

\title{Photon lifetime in a cavity containing a slow-light medium}


\author{T.~Laupr\^etre$^{1,*}$, C.~Proux$^1$, R.~Ghosh$^2$, S.~Schwartz$^3$, F.~Goldfarb$^1$, and F.~Bretenaker$^{1}$}

\address{
$^1$Laboratoire Aim\'e Cotton, CNRS-Universit\'e Paris Sud 11, 91405 Orsay Cedex, France
\\
$^2$School of Physical Sciences, Jawaharlal Nehru University, New Delhi 110067, India
\\
$^3$Thales Research and Technology France, Campus Polytechnique, 91767 Palaiseau Cedex, France
\\
$^*$Corresponding author: thomas.laupretre@lac.u-psud.fr
}

\begin{abstract}
We investigate experimentally the lifetime of the photons in a cavity containing a medium exhibiting strong positive dispersion. This intracavity positive dispersion is provided by a metastable helium gas at room temperature in the electromagnetically induced transparency (EIT) regime, in which light propagates at a group velocity of the order of $10^4$~m.s$^{-1}$. The results definitely prove that the lifetime of the cavity photons is governed by the group velocity of light in the cavity, and not its phase velocity.
\end{abstract}

] 

A lot of recent research has been dedicated to the dramatic changes in the absorption and dispersion properties of optical media induced by coherent processes \cite{Boller1991,harris1992,boyd2003}. EIT can, for example, greatly reduce the group velocity of light even in a gas at room temperature \cite{kash1999}. Even negative dispersion can be achieved through EIT  \cite{mikhailov2004,goldfarb2009}, as well as through other phenomena such as bi-frequency Raman gain \cite{dogariu2001}, leading to supraluminal light or negative group velocity. Besides, the improvement of the sensitivity of sensors is also an active field of investigation. Among the sensors, optical cavities are often good candidates. One approach to enhance their sensitivity is to insert a highly dispersive medium inside the cavity. For example, coherent population trapping and EIT have been used to reduce the linewidth of a resonator \cite{Muller1997,Lukin1998,Wang2000}. Moreover, it has been argued that the scale factor of a ring laser gyro can be reduced or enhanced depending on  the presence of a positive or negative dispersion medium inside the cavity \cite{Shahriar2007PRA,Pati2007PRL}. In several papers, such proposals are accompanied by the statement that the phase velocity of light rules the cavity decay time $\tau_{\mathrm{cav}}$ \cite{Shahriar2007PRA,Salit2007,Savchenkov2006}. But others simply assume that $\tau_{\mathrm{cav}}$ is rather connected with the group velocity \cite{vanExter1991,Joshi2004}. In this paper we investigate experimentally the relationship between the decay time of a cavity filled with a positive dispersive medium and the reduced group velocity.

\begin{figure}[]
\begin{center}
\includegraphics[width=8.5 cm]{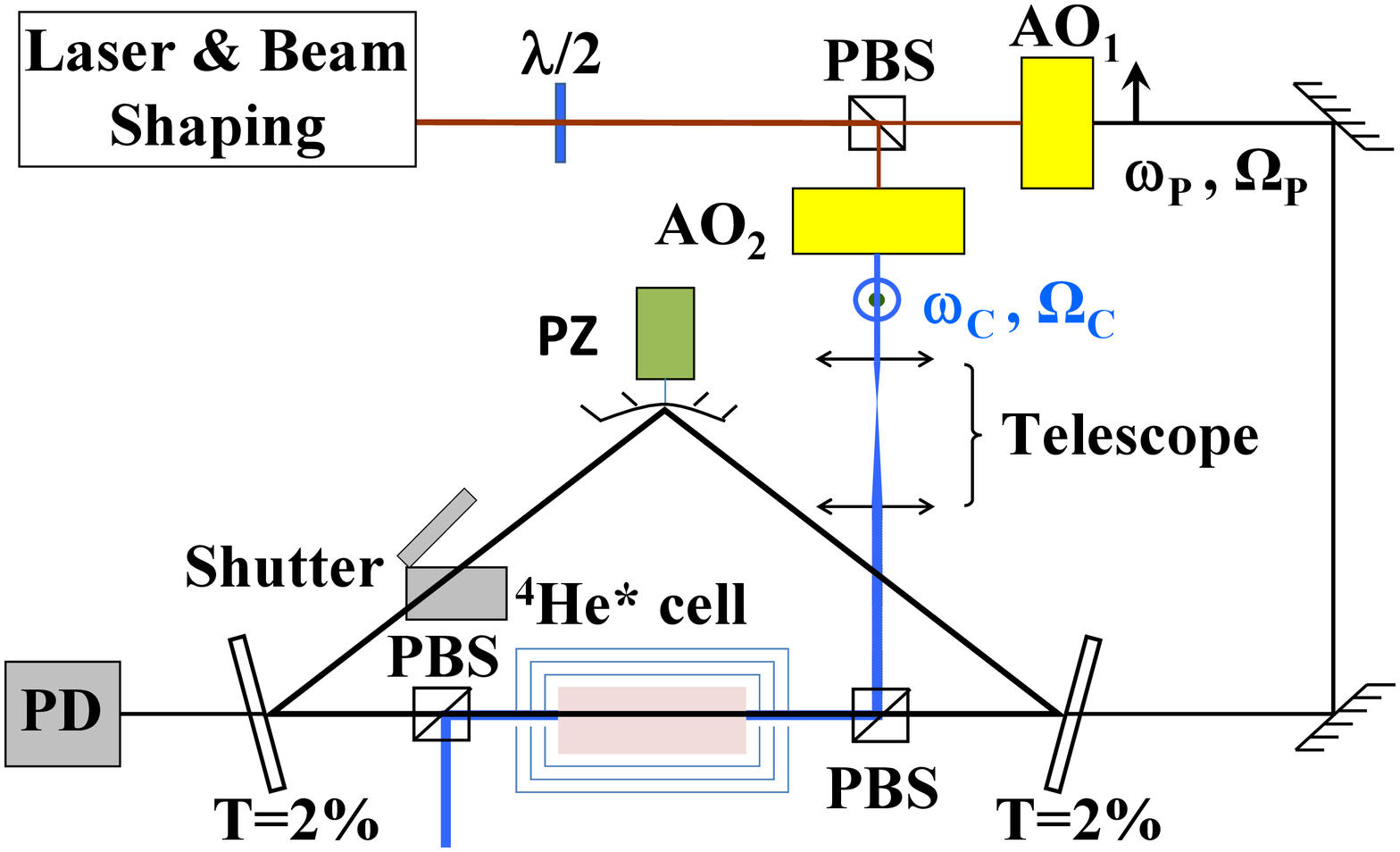}
\end{center}
\caption{Experimental setup. PBS: Polarization Beam Splitter, AO$_{1,2}$: Acousto-Optic modulators, PD: Photodetector, PZ: Piezoelectric actuator. The L = 2.4 m ring cavity is resonant for the probe field ($\omega_\textrm{P}$, $\Omega_\textrm{P}$).}
\label{fig:Fig01}
\end{figure}

To this aim, we use a 2.4-m long ring cavity made of two plane mirrors with 2\% transmission and a high reflectivity concave mirror (5-m radius of curvature) mounted on a piezoelectric actuator  (see Fig.\;\ref{fig:Fig01}). The dispersive medium is a 6\,cm long cell filled with 1\,Torr of helium at room temperature. Some helium atoms are excited in the $^3$S$_1$ metastable state with an RF discharge at 27\,MHz. The cell is located inside a $\mu$-metal shield to avoid magnetic disturbances. Metastable helium is well known to exhibit a pure three-level $\Lambda$ system when excited at 1.083\,$\mu$m between the $2^3$S$_1$ and $2^3$P$_1$ energy levels using circularly polarized light. In these conditions, narrow EIT windows can be obtained \cite{goldfarb2008}. Light at 1.083\,$\mu$m is provided by a single-frequency diode laser. The frequencies $\omega_{\textrm{C,P}}$ and Rabi frequencies $\Omega_{\textrm{C,P}}$ of our coupling and probe beams are driven by two acousto-optic modulators (AOs). A telescope expands the coupling beam diameter up to 1~cm, which is much larger than the probe beam diameter. The cavity is resonant only for the probe field as two polarization beam splitters drive the coupling beam inside and outside the cavity (see Fig. \ref{fig:Fig01}). Notice here that we use two linearly polarized fields for reasons of experimental simplicity. Indeed, we have shown \cite{Laupretre2009} that use of linear polarizations instead of circular polarizations corresponds to just a change of basis for the ground state subspace but does not alter the characteristics of the observed EIT.

\begin{figure}[]
\begin{center}
\includegraphics[width=8.5 cm]{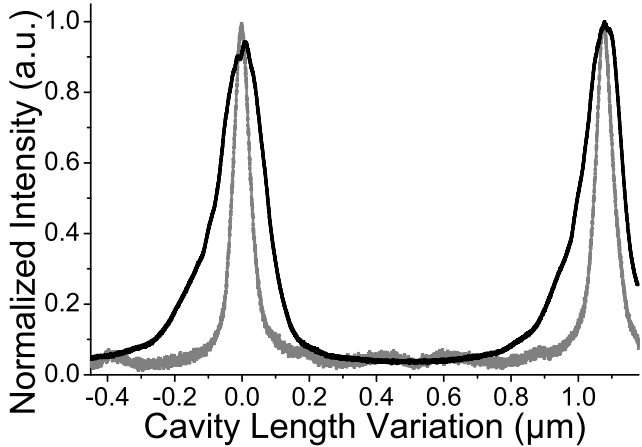}
\end{center}
\caption{Experimental scan of the cavity transmission without the discharge on the atoms (gray line) and with the discharge and a 9~mW coupling power (black line, intracavity probe power about 0.6~mW). The respective measured finesses are 30 and 7.5.}
\label{fig:Fig02}
\end{figure}
If we want to determine whether the cavity decay rate $1/\tau_{\mathrm{cav}}$ is ruled by the phase or group velocity, we need an estimate of the cavity round-trip losses. With this aim, we first probe the transmission of the cavity when its length is scanned using the piezoelectric actuator acting on the concave mirror. A typical result is reproduced as a black line in Fig.\;\ref{fig:Fig02} when a coupling beam (9~mW), tuned at the center of the transition Doppler profile, is applied to the atoms. The probe beam frequency is tuned at the maximum of the EIT transmission window, where the light slowing down effect is maximum. This result shows that our resonator indeed behaves like a regular optical cavity and will be used later to extract a measurement of the intracavity losses. For the sake of comparison, we reproduce as a grey line in Fig.\;\ref{fig:Fig02} the transmission of the cavity without the discharge. The finesse is equal to 30, corresponding to about 20\;\% losses per round-trip which are due to the mirror transmissions and the uncoated intracavity cell windows and polarizing beam splitters. Note that with the discharge on and in the absence of coupling beam, the strong cell absorption (about 97\;\%) prevents us from showing a cavity transmission profile. The cavity behavior can then be restored by turning on the coupling field and the finesse increases with the coupling field power. The fact that the finesse of the cavity in the presence of EIT (black curve of Fig.\;\ref{fig:Fig02}) is smaller than without the atoms is due to the residual absorption of the atoms \cite{goldfarb2008,Laupretre2009}.

\begin{figure}[]
\begin{center}
\includegraphics[width=8.5 cm]{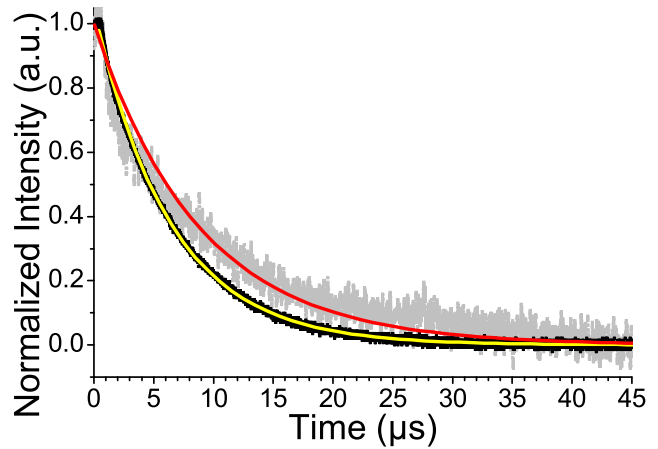}
\end{center}
\caption{Experimentally measured decays of the cavity intensity in the presence of intracavity slow light for coupling powers of 3~mW (gray) and 9~mW (black). The probe power incident on the cavity is 1.45~mW. The thin full lines are fits by exponential decays, leading to $\tau_{\mathrm{cav}}=8.8\;\mu$s and $\tau_{\mathrm{cav}}=6.3\;\mu$s, respectively.}
\label{fig:Fig03}
\end{figure}

\begin{figure}[]
\begin{center}
\includegraphics[width=8.5 cm]{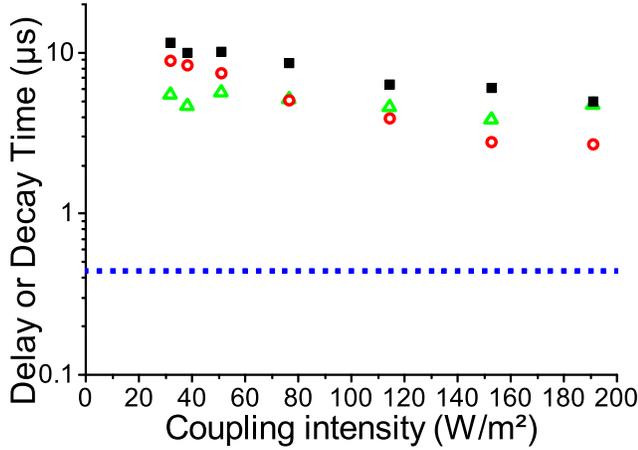}
\end{center}
\caption{Evolution of the delay and decay time versus coupling power. Black squares: measured cavity decay time $\tau_{\mathrm{cav}}$ with intracavity slow light. Open red circles: measured group delay $\tau_{\mathrm{g}}$ through the EIT medium only (shutter closed). Open green triangles: cavity decay time calculated from measured $\tau_{\mathrm{g}}$ and cavity round-trip transmission $\mathcal{T}$. Dotted line: detection response time.}
\label{fig:Fig04}
\end{figure}

In order to measure $\tau_{\mathrm{cav}}$ for this cavity, one slowly scans its length and turns off the probe beam using AO$_1$ when it reaches resonance. One then records the decay of the cavity output power while the coupling beam is kept on. Fig.\,\ref{fig:Fig03} shows the decay of the intensity at the output of the cavity for two different coupling powers. These signals can be well fitted by exponential decays. The time constants of these fits provide the measured lifetime $\tau_{\mathrm{cav}}$ of the photons in the cavity.
The black squares in Fig.~\ref{fig:Fig04} represent the measurements of $\tau_{\mathrm{cav}}$ for different values of the coupling beam power. We notice that the recorded lifetimes are of the order of several microseconds, which is at least two orders of magnitude longer than what would be expected if $\tau_{\mathrm{cav}}$ were governed by the phase velocity of light. Indeed, when we switch off the RF discharge, i.e., when the helium atoms are back in the ground state and transparent, we observe that the cavity decay time is so short that it cannot be measured with our detection: the measured detection response time is about 450\,ns, represented by the blue dotted line in Fig.~\ref{fig:Fig04}.

To check the consistency of our measurements, we measure the group delay $\tau_{\mathrm{g}}$ through the cell in the EIT regime. For this, we just introduce a shutter inside the cavity, as shown in Fig.~\ref{fig:Fig01}, and observe the propagation of a sinusoidal probe (intensity modulated at 2~kHz) through the cell sandwiched between the two plane mirrors. By comparing the modulation phase shift with and without EIT, we obtain the measurements of $\tau_{\mathrm{g}}$ depicted by the red open circles in Fig.~\ref{fig:Fig04}. As expected, the values of $\tau_{\mathrm{cav}}$ are larger than our measurements of $\tau_{\mathrm{g}}$, showing that the group velocity governs the photon lifetime in this cavity.

We can try and go further into our data analysis by linking our measurements of $\tau_{\mathrm{g}}$, $\tau_{\mathrm{cav}}$, and the cavity losses. Indeed, if we assume that the photon lifetime is driven by the group velocity, we have
 \begin{equation}
\tau_{\mathrm{cav}} = -\frac{\tau_{\mathrm{g}}}{\ln(\mathcal{T})}, \label{eq1}
\end{equation}
where $\mathcal{T}$ is the intensity transmission for one round-trip inside the cavity. It is the product of the reflectivities of the three mirrors and of the transmission of the cell which increases with the coupling intensity. $\mathcal{T}$ is experimentally derived from the width of the transmission peaks similar to those in Fig.~\ref{fig:Fig02}. Using Eq.~(\ref{eq1}), we obtain the green open triangles of Fig.~\ref{fig:Fig04}. We first notice that the measurements of $\tau_{\mathrm{cav}}$ (black squares) and its determination using Eq.~(\ref{eq1}) yield the same order of magnitude, reinforcing our  conclusion that the cavity decay time is governed by the group velocity of light. The small difference between the two sets of measurements has several origins: i) Eq.~(\ref{eq1}) is no longer strictly valid when $\mathcal{T}$ becomes small due to the poor transmission of the cell occurring for the smallest values of the coupling intensity \cite{goldfarb2008}; ii) the measurements of $\tau_{\mathrm{g}}$ are performed with a  1.45~mW power incident on the cavity input mirror, i.e., with about 30~$\mu$W incident on the cell, while the measurements of $\tau_{\mathrm{cav}}$ are performed when the cavity is resonant, i.e., with much more power saturating the EIT transmission of the cell; iii) when the cavity is resonant, the radius of the probe beam is determined by the cavity eigenmode, while when we measure $\tau_{\mathrm{g}}$, the beam is directly incident on the cell; iv) the measurements of the cavity finesse using signals similar to the one in Fig.~\ref{fig:Fig02} lead to an underestimation of $\mathcal{T}$ because the intracavity probe power is larger at exact resonance than on the wings of the resonance. Thus, the EIT saturation varies inside the cavity resonance and with the coupling power, making the complete quantitative comparison of the measurements of $\tau_{\mathrm{cav}}$ on the one hand, and $\tau_{\mathrm{g}}$ and $\mathcal{T}$ on the other hand, impossible.

In conclusion, we have shown experimentally that the introduction of a positive dispersive medium inside an optical cavity increases the photon lifetime by several orders of magnitude. This definitively proves that the cavity decay rate is driven by the intracavity group velocity of light, as long as the considered light frequency lies within the bandwidth of the phenomenon inducing slow light. This raises the question of whether the frequency noise of a laser containing such a slow or fast light medium depends on the phase or group velocity of light in this medium. The answer to this may have important consequences for ring laser gyros \cite{dorschner1980}. Future investigations comprise of measuring the lifetime of the photons in our cavity in the presence of negative dispersion leading either to superluminal propagation or negative group velocity  \cite{goldfarb2009}, and the experimental investigation of the noise of a laser containing such a medium.

\section*{Acknowledgments}
RG wishes to thank Universit\'e Paris Sud 11 for an Invited Professor position during June 2010. This work was partially supported by the Indo-French Center for the Promotion of Advanced Research.


\begin{thebibliography}{99}

\bibitem{Boller1991}
K.-J. Boller, A.~Imamoglu, and S.~E. Harris, Phys. Rev. Lett. \textbf{66},
  2593 (1991).

\bibitem{harris1992}
S.~E. Harris, J.~E. Field, and A.~Kasapi, Phys. Rev. A \textbf{46}, R29
  (1992).

\bibitem{boyd2003}
M.~S. Bigelow, N.~N. Lepeshkin, and R.~W. Boyd, Science \textbf{301},  200 (2003).

\bibitem{kash1999}
M.~M. Kash, V.~A. Sautenkov, A.~S. Zibrov, L.~Hollberg, G.~R. Welch, M.~D.  Lukin, Y.~Rostovtsev, E.~S. Fry, and M.~O. Scully, Phys. Rev. Lett. \textbf{82}, 5229 (1999).

\bibitem{mikhailov2004}
E.~E. Mikhailov, V.~A. Sautenkov, I.~Novikova, and G.~R. Welch, Phys. Rev. A \textbf{69}, 063808 (2004).

\bibitem{goldfarb2009}
F.~Goldfarb, T.~Laupr\^{e}tre, J.~Ruggiero, F.~Bretenaker, J.~Ghosh, and  R.~Ghosh, C. R. Physique \textbf{10}, 919 (2009).

\bibitem{dogariu2001}
A.~Dogariu, A.~Kuzmich, and L.~J. Wang, Phys. Rev. A \textbf{63}, 053806 (2001).

\bibitem{Muller1997}
G.~M\"uller, M.~M\"uller, A.~Wicht, R.-H. Rinkleff, and K.~Danzmann, Phys. Rev. A  \textbf{56}, 2385 (1997).

\bibitem{Lukin1998}
M.~D. Lukin, M.~Fleischhauer, M.~O. Scully, and V.~L. Velichansky, Opt. Lett.  \textbf{23}, 295 (1998).

\bibitem{Wang2000}
H.~Wang, D.~J. Goorskey, W.~H. Burkett, and M.~Xiao, Opt. Lett.  \textbf{25}, 1732 (2000).

\bibitem{Shahriar2007PRA}
M.~S. Shahriar, G.~S. Pati, R.~Tripathi, V.~Gopal, M.~Messall, and K.~Salit, Phys. Rev. A \textbf{75}, 053807 (2007).

\bibitem{Pati2007PRL}
G.~S. Pati, M.~Salit, K.~Salit, and M.~S. Shahriar, Phys. Rev. Lett. \textbf{99}, 133601 (2007).

\bibitem{Salit2007}
M.~Salit, G.~S. Pati, K.~Salit, and M.~S. Shahriar,  J. Mod. Opt. \textbf{54}, 2425 (2007).

\bibitem{Savchenkov2006}
A.~A. Savchenkov, A.~B. Matsko, and L.~Maleki, Opt. Lett. \textbf{31}, 92 (2006).

\bibitem{vanExter1991}
M.~P. van Exter, W.~A. Hamel, and J.~P. Woerdman, Phys. Rev. A \textbf{43}, 6241 (1991).

\bibitem{Joshi2004}
A.~Joshi and M.~Xiao, in \emph{Conference on Laser and Electro-Optics (CLEO)},  (Optical Society of America, 2004), paper IMP3.

\bibitem{goldfarb2008}
F.~Goldfarb, J.~Ghosh, M.~David, J.~Ruggiero, T.~Chaneli\`ere, J.-L. {Le  Gou\"et}, H.~Gilles, R.~Ghosh, and F.~Bretenaker, Europhys. Lett.  \textbf{82}, 54002 (2008).

\bibitem{Laupretre2009}
T.~Laupr\^{e}tre, J.~Ruggiero, R.~Ghosh, F.~Bretenaker, and F.~Goldfarb, Opt. Express \textbf{17},  19444 (2009).

\bibitem{dorschner1980}
T.~A. Dorschner, H.~A. Haus, M.~Holz, I.~W. Smith, and H.~Statz, IEEE J. Quantum Electron. \textbf{16}, 1376  (1980).

\end{thebibliography}
\end{document}